\documentclass[twocolumn]{revtex4}
\usepackage{graphicx}
\usepackage{courier}
\usepackage{amssymb,amsmath}
\usepackage{longtable} 
\begin{document}

\title{Turbulent magnetic field amplification from the smallest to the largest magnetic Prandtl numbers}
\author{S. Bovino\footnote{sbovino@astro.physik.uni-goettingen.de}$^1$, D. R. G. Schleicher$^1$\footnote{dschleich@astro.physik.uni-goettingen.de}, and J. Schober$^2$\footnote{Schober@stud.uni-heidelberg.de}}
\affiliation{$^1$Institut f\"ur Astrophysik, Georg-August-Universit\"at, Friedrich-Hund Platz 1, D-37077, G\"ottingen, Germany\\
$^2$Institut f\"ur Theoretische Astrophysik, Zentrum f\"ur Astronomie der Universit\"at Heidelberg, Albert-Ueberle Strasse, 2, D-69120 Heidelberg, Germany}
\begin{abstract}
The small-scale dynamo provides a highly efficient mechanism for the conversion of turbulent into magnetic energy. In astrophysical environments, such turbulence often occurs at high Mach numbers, implying steep slopes in the turbulent spectra. It is thus a central question whether the small-scale dynamo can amplify magnetic fields in the interstellar or intergalactic media, where such Mach numbers occur. To address this long-standing issue, we employ the Kazantsev model for turbulent magnetic field amplification, systematically exploring the effect of different turbulent slopes, as expected for Kolmogorov, Burgers, the Larson laws and results derived from numerical simulations. With the framework employed here, we give the first solution encompassing the complete range of magnetic Prandtl numbers, including Pm $\ll$ 1, Pm $\sim$ 1 and Pm $\gg$ 1.

We derive scaling laws of the growth rate as a function of hydrodynamic and magnetic Reynolds number for Pm $\ll1$ and Pm $\gg1$ for all types of turbulence. 
A central result concerns the regime of Pm $\sim1$, where the magnetic field amplification 
rate increases rapidly as a function of Pm. This phenomenon occurs for all types of turbulence we explored. We further find that the dynamo growth rate can be decreased by a few orders of magnitude for turbulence spectra steeper than Kolmogorov. We calculate the critical magnetic Reynolds number Rm$\rm_c$ for magnetic field amplification, 
which is highest for the Burgers case. As expected, our calculation shows a linear behaviour of the amplification rate close to the threshold proportional to $(\rm Rm - Rm_c$).  
Based on the Kazantsev model, we therefore expect the existence of the small-scale dynamo for any given value of Pm, as long as the magnetic Reynolds number is above the critical threshold.\end{abstract}

\maketitle
\section{Introduction}
The conversion of kinetic energy into magnetic energy, the so-called dynamo action, plays an important role in a wide range of astrophysical applications. These include the amplification of 
magnetic fields in the surface of the Sun \cite{Graham2009,Graham2010}, the origin of magnetic fields in galaxies \citep{Beck96, Arshakian09,deSouza10}, galaxy clusters \cite{SubramanianS2006,Shukurov2006}, 
the large-scale structure of the Universe \cite{Ryu08} and even the formation of the first stars and galaxies \cite{Schleicher2010, Sur2010, Turk12, Federrath2011ApJ}. 

Depending on the environment, the turbulent Mach number $\mathcal{M}$, i.e. the ratio of the turbulent velocity $v_t$ to the sound speed $c_s$, may  vary considerably. In stellar interiors and during primordial star formation, 
we expect subsonic turbulence close to the well-studied case of Kolmogorov \cite{Kolmogorov1941}, while turbulence in galaxies may exhibit Mach numbers considerably larger than 1, which is typical for Burgers-type 
turbulence \cite{Burgers1948}. Observations in turbulent molecular clouds revealed the so-called Larson laws, hinting towards an intermediate case in between Burgers and Kolmogorov turbulence \cite{Larson81}.
 The Larson laws consist of: (i) a power relationship between the velocity dispersion $\sigma_v$, and the spatial scale of the emitting volume L, $\sigma_v \propto$ L$^{0.38}$, (ii) selfgravitational equilibrium, 2$\sigma_v$L$^2$/GM $\propto$ 1, and (iii) an inverse relationship between the mean density n, and size of the cloud, n $\propto$ L$^{−1.1}$. The Larson laws therefore describe observed cloud properties in the turbulent and highly compressible interstellar medium.
 
 For the modeling of turbulent magnetic field amplification, the well-studied Kraichnan-Kazantsev model \cite{Kazantsev1968,Kraichnan1968,Kraichnan1994} is often employed. The Kraichnan model \cite{Kraichnan1968,Kraichnan1994} considers the advection of a passive scalar field, which may for instance represent some chemical species, while the Kazantsev model \cite{Kazantsev1968} describes the turbulent diffusion of a passive vector field. Both models employ the same assumptions, in particular a velocity field based on a zero mean Gaussian random process as well as a $\delta$-correlation in time. Both models have been largely used over the years and applied to diverse physical circumstances (e.g. \cite{Schober2012,Subramanian1997,Brandenburg20051,Boldyrev2004,Boldyrev2005,Malyshkin2010} and \cite{Chertkov1999,Falkovich2001}) and represent an important theoretical laboratory for studying the multiparticle statistics in fluid turbulence. As we are in the following considering the amplification of vector quantities, we will adopt the term Kazantsev model for brevity. 

The classical studies of the Kazantsev model typically focused only on Kolmogorov turbulence. As discussed above, the latter is often not applicable in astrophysical environments, and steeper slopes for the turbulent spectra are frequently found both in numerical simulations and observational data sets. A systematic exploration of such steeper slopes is therefore in order. We pursue that here for the full range of magnetic Prandtl numbers Pm, denoting the ratio of kinetic viscosity $\nu$ and 
magnetic viscosity $\eta$, from Pm $\ll$ 1 to Pm $\gg$ 1, covering the complete range that may occur in astrophysical environments. For instance, in stellar interiors, Pm is considerably smaller than 1, while the gas in galaxies and in the intergalactic medium exhibits magnetic Prandtl numbers much larger than 1. In primordial clouds, the latter can also change during the evolution as a result of ambipolar diffusion \cite{Schober2012a}. Over the last years, studies on both very large magnetic Prandtl numbers \cite{Schober2012,Schleicher2009,Schleicher2010,Haugen2004,Schekochihin2005,Subramanian1997}, where the amplification is caused  by random stretching of the magnetic field at the viscous scale, and very small magnetic Prandtl numbers \cite{Boldyrev2004,Kleeorin2012,Iskakov2007,Malyshkin2010,Scheko07}, where it is driven by the inertial-range velocity field,  have been reported. The above-mentioned studies employed a large range of analytical and numerical methods, which usually are only applicable in a limited regime. Analytical studies making use of the Kazantsev model may for instance treat the regime of Pm~$\ll$~1 and Pm~$\gg$~1, but not the regime of Pm~$\sim$~1. 
Conversely, numerical simulations are bound to the regime Pm $\sim$ 1 \cite{Plunian2010,Sahoo2011,Ponty2005}. A comparison of these approaches is thus very difficult. As mentioned above, most studies so far explored only the effect of Kolmogorov turbulence, 
although the interstellar medium in the first galaxies is expected to be in a highly compressible regime and potentially closer to Burgers turbulence.

In this paper, we accurately solve (via numerical integration) the Kazantsev equation and calculate the growth rate of the small-scale dynamo for the complete range of magnetic Prandtl numbers, from the smallest to the largest, for different types of turbulence. We define the kinematic and magnetic Reynolds numbers as Re=VL/$\nu$ and Rm=VL/$\eta$, where V is the typical velocity at the largest scale of the inertial range L. We thus have Pm = Rm/Re. We note that, in addition to microphysical processes like Ohmic diffusion, turbulent diffusion process may play a substantial role, as pointed out by \citet{Weinan2001,Sreenivasan2010}. Here, these effects are incorporated into our model in terms of a diffusion coefficient. Covering the range sketched above provides solutions for the growth rate in the astrophysically relevant regimes, 
and simultaneously establishes a pathway for a comparison with 3D magneto-hydrodynamical simulations. In the following Section, we present the Kazantsev model and our numerical treatment. Subsequently, we show the 
different growing modes for Kolmogorov turbulence, discuss the dependence on different types of turbulence, analyze the behavior close to the threshold and discuss the implications of our results.
 \section{Kazantsev model and numerical solution}
The magnetic field amplification is governed by the induction equation
\begin{equation}\label{eq:induction}
	\frac{\partial {\bf B}}{\partial t} = \nabla\times {\bf v}\times {\bf B} - \eta\nabla\times\nabla\times {\bf B},
\end{equation}
where {\bf B} is the magnetic field, and {\bf v} is the fluid velocity. 

The Kazantsev model \cite{Kazantsev1968} describes the small-scale dynamo process for a Gaussian random velocity field which is $\delta$-correlated in time. For this purpose, we introduce the covariance tensor as
\begin{equation}\label{eq:tensor}
	\langle\delta v_i({\bf r_1}, t)\delta v_j({\bf r_2}, s)\rangle = T_{ij}(r)\delta(t-s)
\end{equation} 
being $T_{ij}(r)$ the two-point correlation function, defined by Batchelor \citep{Batchelor1950} as 
\begin{equation}\label{eq:correlation}
	T_{ij}(r) = \left(\delta_{ij}-\frac{r_ir_j}{r^2}\right)T_N(r) + \frac{r_ir_j}{r^2}T_L(r)
\end{equation}
with $T_N(r)$ and $T_L(r)$ being the transversal and longitudinal part, respectively.
We use the model for the correlation function of the turbulent velocity field has presented in \citet{Schober2012}
\begin{equation}
  T_\text{L}(r) = \begin{cases}
               \frac{VL}{3}\left(1-Re^{(1-\vartheta)/(1+\vartheta)}\left(\frac{r}{L}\right)^{2}\right) & 0<r<\ell_\nu \\
               \frac{VL}{3}\left(1-\left(\frac{r}{L}\right)^{\vartheta+1}\right)                   & \ell_\nu<r<L \\
               0                                                                               &  L<r,
            \end{cases}
\end{equation}
where $\ell_\nu=L~Re^{-1/(\vartheta+1)}$ denotes the cutoff scale of the turbulence, i.~e.~the viscous scale, and $L$ the length of the largest eddies.\\
The transverse correlation function for the general slope of the turbulent velocity spectrum is
\begin{equation}
  T_\text{N}(r) = \begin{cases}
                     \frac{VL}{3}\left(1-t(\vartheta)Re^{(1-\vartheta)/(1+\vartheta)} \left(\frac{r}{L}\right)^{2}\right)      & 0<r<\ell_\nu \\
               \frac{VL}{3}\left(1-t(\vartheta)\left(\frac{r}{L}\right)^{\vartheta+1}\right)                              & \ell_\nu<r<L \\
               0                                                                                                       & L<r,
                \end{cases}
\end{equation}
with $t(\vartheta)=(21-38\vartheta)/5$. 
In this formulation we neglect the effect of the helicity which is not important at small scales. We note that, when refering to different types of turbulence below, we refer to the velocity correlation functions defined here, adopting the respective slope of the turbulence model, $\delta v \propto \ell^\vartheta$. These models of course only represent an approximation of real astrophysical turbulence, where the correlation time is finite.

By analogy, one can define the components $M_N$ and $M_L$ for the correlation function of B, where $M_L$ and $M_N$ represent the correlations between the same components  (e.g. $\langle B_x(0) B_x(r)\rangle$) and between different components (e.g. $\langle B_x(0)B_y(r)\rangle$) of {\bf B}. In order to ensure the constraint $\nabla\cdot\vec{B}=0$, these functions need to obey the following equation:\begin{equation}
M_N=\frac{1}{2r}\frac{\partial }{\partial r}\left(r^2 M_L(r)\right).
\end{equation}

Solutions for the Kazantsev equation have been previously proposed in many works and for different parameters by assuming isotropy and homogeneity \cite{Subramanian1997,Boldyrev2005}. In a recent study, 
Schober et al. \cite{Schober2012} discussed an analytical solution based on the quantum-mechanical WKB approximation for the case of infinite magnetic Prandtl numbers for different types of turbulence. 
By using the ansatz given in their paper, it is possible to obtain the Kazantsev equation from Eq.~(\ref{eq:induction}) as
\begin{equation}\label{eq:radial}
	-\kappa(r)\frac{d^2\Psi(r)}{dr^2} + U(r)\Psi(r) = -\Gamma\Psi(r),
\end{equation}
with $\Psi(r)$ related to $M_L$ through the formula $M_L(r,t)=~\Psi(r)e^{2\Gamma t}/(r^2\sqrt\kappa) $.
Equation (\ref{eq:radial}) formally looks like the quantum mechanical Schr\"odinger equation with $\Gamma$ being the growth rate and $U(r)$ the potential, defined as
\begin{equation}\label{eq:potential}
	U(r) = \frac{\kappa''}{2} -\frac{(\kappa')^2}{4\kappa} + \frac{2T'_N}{r} + \frac{2(T_L - T_N - \kappa)}{r^2}.
\end{equation}
The diffusion of the magnetic correlation $\kappa(r)$ is defined as follow:
\begin{equation}\label{eq:diffusion}
	\kappa(r) = \eta + T_L(0) - T_L(r).
\end{equation}
The latter contains the magnetic diffusivity $\eta$ and the scale-dependent turbulent diffusion $T_L(0) - T_L(r)$ (see also \citet{Weinan2001,Sreenivasan2010} for the implications of turbulent diffusion in the Kraichnan model).
It is worth noting that numerical simulations show the presence of shear even at high Mach numbers, implying that such turbulent diffusion will be available both in the compressible and the incompressible 
case \cite{Federrath2010,Federrath2011}.
Note also that a solution $\Gamma>0$ for Eq. (\ref{eq:radial}) exists only for $U(r)$ sufficiently negative in some region of $r$. The problem is then reduced to a search of bound states for $U(r)$.

Defining $x = \ln(r)$ and $\Psi(x) = e^{x/2}\theta(x)$, the Kazantsev equation reads
\begin{equation}\label{eq:katzantsev}
	\frac{d^2\theta(x)}{dx^2} + p(x)\theta(x) = 0
\end{equation}
with 
\begin{equation}\label{eq:pfunction}
	p(x) = -\frac{(\Gamma + U(x))e^{2x}}{\kappa(x)}-\frac{1}{4}.
\end{equation}
It is useful to introduce the normalised growth rate
\begin{equation}
	\bar{\Gamma} = \frac{L}{V}\Gamma, 
\end{equation}
with $V$ and $L$ the turbulent velocity and the length of the largest eddies, respectively. For further details on the derivation of Eq.s (\ref{eq:katzantsev}) and (\ref{eq:pfunction}) we refer to Schober et al. \cite{Schober2012}.

We solve  Eq. (\ref{eq:katzantsev}) employing the Numerov algorithm \cite{Numerov1924} which is well-suited for the study of second order problems which contain no first order derivative.
Two boundary conditions for $\theta(x)$ are needed in order to obtain the solution. We use
\begin{equation}
  \theta(x)\xrightarrow{x\rightarrow\pm\infty}0. 
\end{equation}
The error in one integration step $h$ is usually $O(h^6$), which leads to a total error in the Numerov 
method of the order $O(h^5)$. We adopt a typical length scale of 1~pc (3.18$\times$10$^{18}$~cm) and a turbulent velocity of $V = 1$~km~s$^{-1}$. All the parameters have been accurately
tested to ensure convergence of $\Gamma$ and of the wavefunction $\theta(x)$.

\section{Growth rate for Kolmogorov turbulence}
\begin{figure}
\includegraphics[width=.45\textwidth]{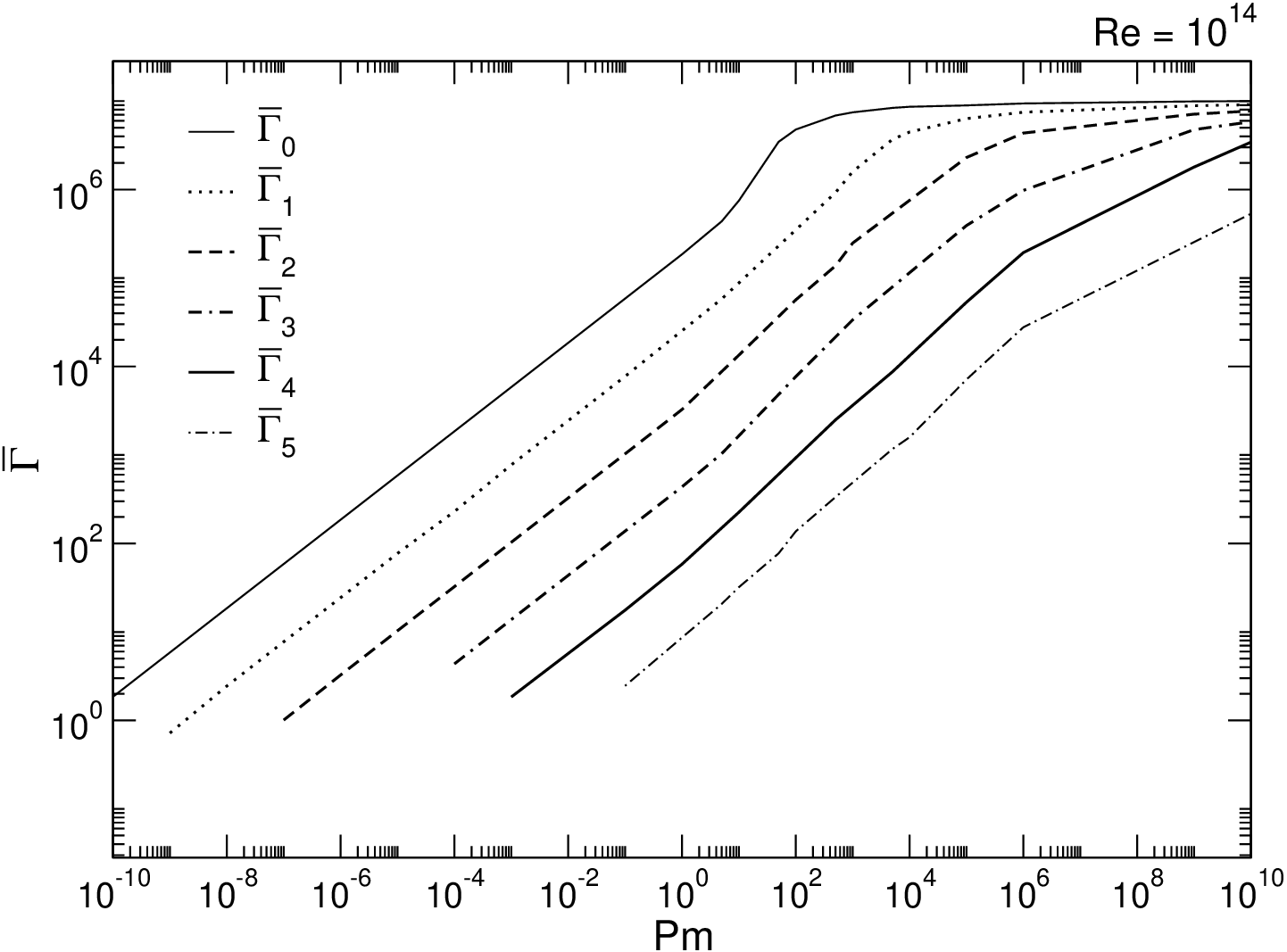}
\caption{Log-Log plot of the computed fastest and slower modes of the normalised growth rate $\bar{\Gamma}$ as a function of $Pm$ for incompressible Kolmogorov turbulence and for Re = 10$^{14}$.}\label{figure1}
\end{figure}

In Fig. \ref{figure1} we report the calculated normalized growth rate $\bar{\Gamma}$ for different modes and Re = 10$^{14}$ for the case of  Kolmogorov turbulence from Pm $\ll1$ up to Pm $\gg1$. For a narrow range of Pm we found a strong increase of the growth rate, in particular for the fastest growing mode, 
which depends on the fact that, for 5 $\le$ Pm $\le$ 10$^{5}$ the potential is negative both in the inertial and in the viscous range yielding two contributions (see Fig. \ref{figure3}). We note that the additional contribution coming from the viscous range is marked only for Pm $>$ 10, even if it starts to appear for Pm $\ge$ 5.  
The higher modes shown in the figure clearly depend on the depth of the potential $U(x)$, and only the main growing mode (the larger in magnitude) was found to exist in the whole range of Pm. 
It is worth noting that the small-scale dynamo amplification can occur also for Pm $\ll$~1. The presence of the higher growing modes is important since it gives an additional contribution to the magnetic field 
amplification that becomes more marked for Pm $\gg$ 1, where the $\bar{\Gamma}$ values are 6 orders of magnitude larger than for Pm $\ll$ 1 (for the same Reynolds number). Especially for Pm $\gg$ 1,  a large number of higher modes has been found. By taking in consideration Burgers turbulence
we found a small number of modes going to a maximum of 3 ($\bar{\Gamma}_0$, $\bar{\Gamma}_1$, and $\bar{\Gamma}_2$) for Pm $\rightarrow\infty$ up to only 1 mode for intermediate and small Pm. This provides a further confirmation of the 
fact that for Kolmogorov turbulence we have a larger amplification of the magnetic field.

\begin{figure}
\includegraphics[width=.45\textwidth]{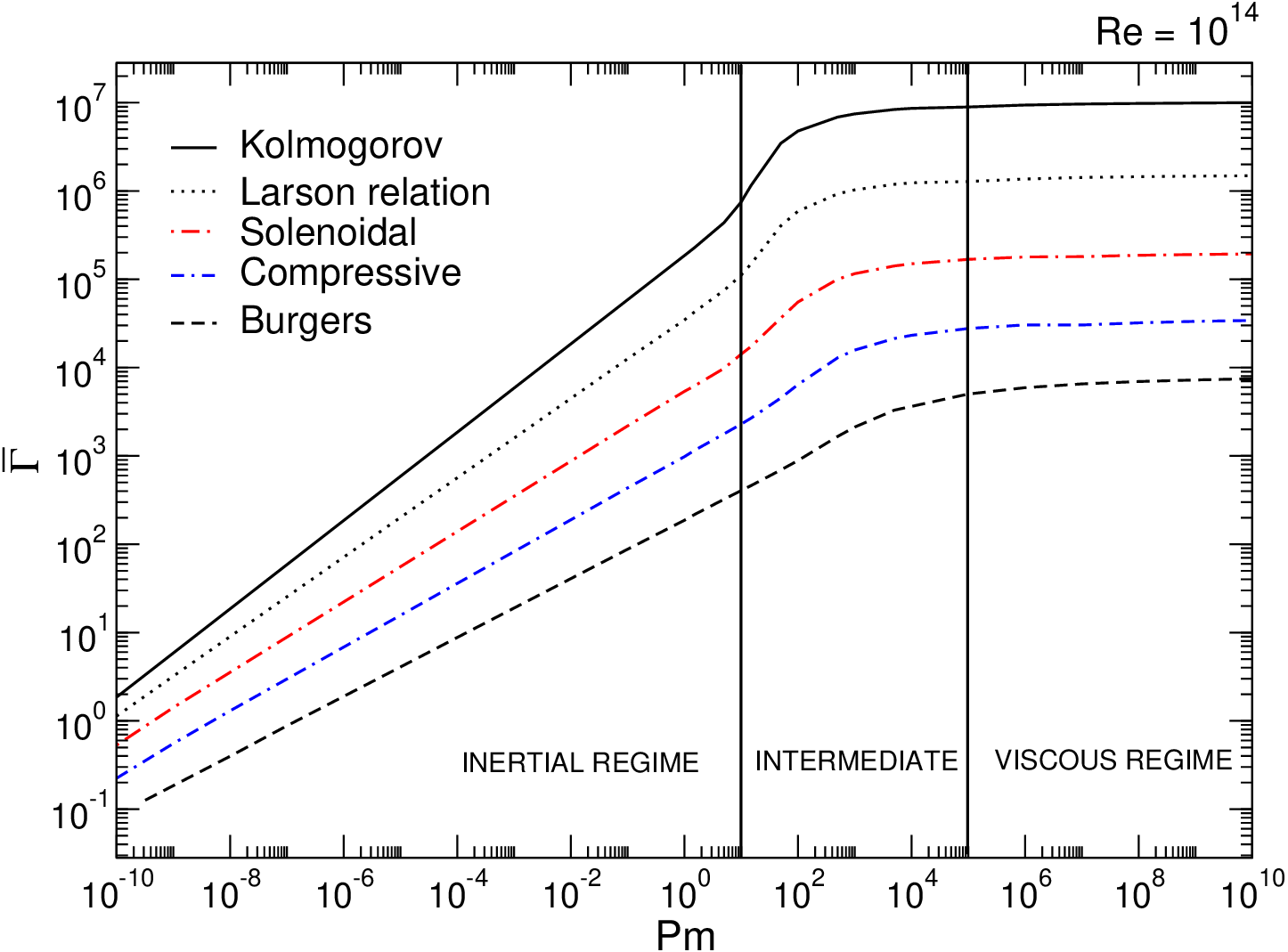}
\caption{Log-Log plot of the computed fastest normalised growth rate $\bar{\Gamma}$ as a function of Pm for five different types of turbulence and for Re = 10$^{14}$. The viscous and inertial regime are explicitely pointed out as well as the intermediate range.}\label{figure2}
\end{figure}

\section{Turbulence dependence of the growth rate}
In Fig. \ref{figure2} we report the fastest growing mode for different types of turbulence indicated by the slope of the turbulent spectrum $\delta v(\ell)\propto \ell^\vartheta$, where $\delta v(\ell)$ is the eddy velocity at the scale $\ell$: Kolmogorov ({\bf $\vartheta=1/3$}), Larson relation ({\bf $\vartheta=0.38$}), solenoidal ({\bf $\vartheta=0.43$}) and compressive forcing  ({\bf $\vartheta=0.48$}) of turbulence, and compressible Burgers ({\bf $\vartheta=1/2$}) . Kolmogorov turbulence 
amplifies the magnetic field more effectively compared to the highly compressible Burgers case. Furthermore, the growth rate linearly increases as Pm$^{1/2}$ for Pm~$\ll$~1 (for Kolmogorov) showing again a flex-point 
at the boundary of the two scales and then becoming constant at very large Pm, consistent with the WKB results \cite{Schober2012}. All the other curves show a similar behaviour.

\begin{figure}
\includegraphics[width=.45\textwidth]{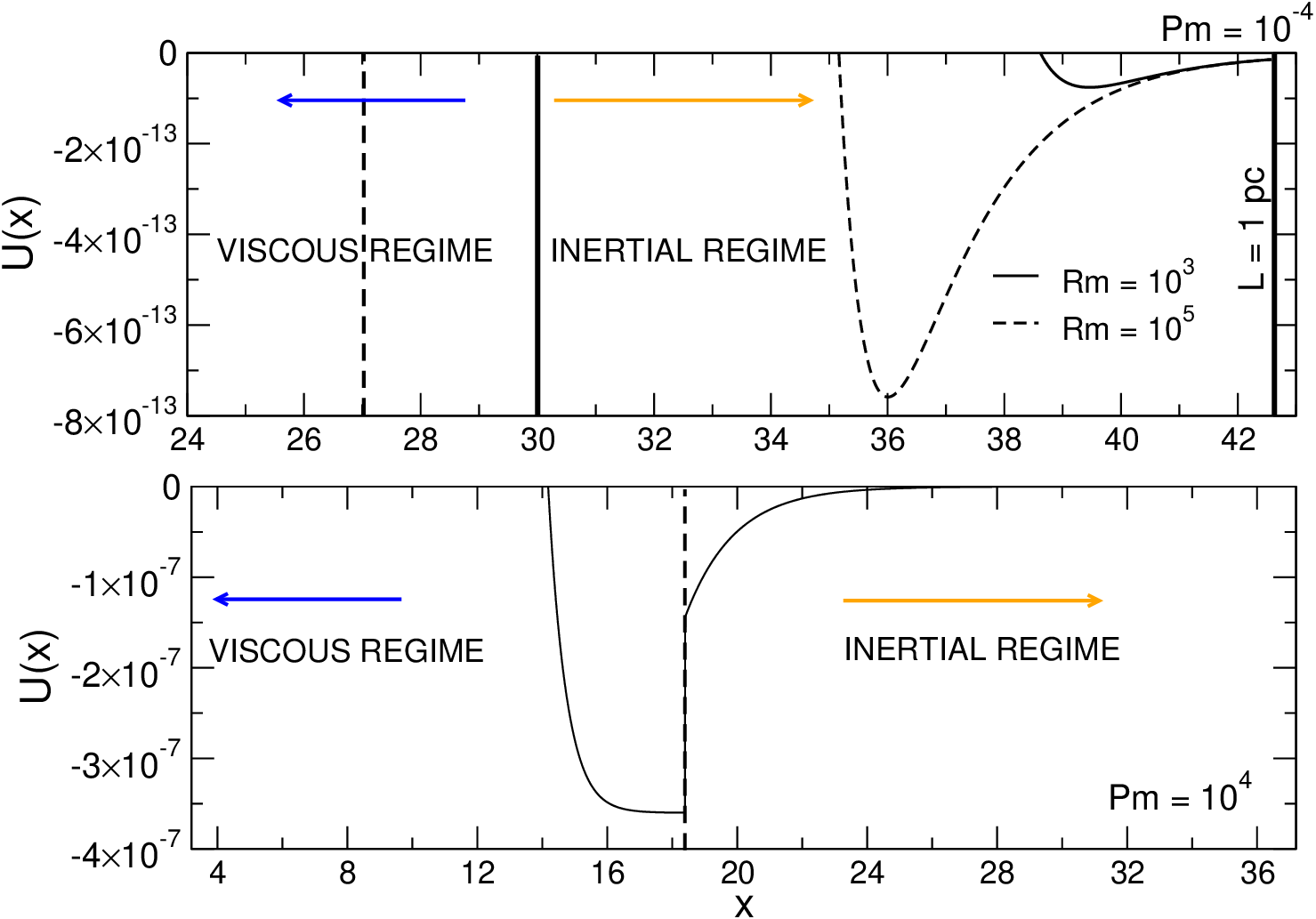}
\caption{The potential $U(x)$ as a function of the coordinate $x$ for two different Pm for Kolmogorov-type turbulence. The upper panel shows Pm=$10^{-4}$, while the bottom panel shows Pm=$10^4$ and Re=$10^{14}$. In both panels the cutoff scale length 
$\ell_\nu$ is shown as a vertical line.}\label{figure3}
\end{figure}

To support  and better explain the results reported in Fig. \ref{figure2} we show in Fig. \ref{figure3} the negative part of the potential $U(x)$ for two different values of Pm and for different magnetic Reynolds numbers. 
On the upper panel it is clearly shown that the potential lies completely in the inertial range and the features are led by the magnetic diffusivity $\eta$.
In the bottom panel the potential in between the two regimes is shown. A discontinuity is present at the length cutoff $\ell_\nu$, reported as a vertical dashed line, and the appearence of a deeper potential in the 
viscous range is the cause of the sudden increase in $\bar{\Gamma}$ for Pm ranging from 10$^1<$ Pm $<10^5$ that we have seen in Fig.s~\ref{figure1} and \ref{figure2}.
Again it is important to note that the contribution coming from the viscous regime exists also for Pm $\sim$ 5, but it is too small and does not affect the integration and the final value of $\bar{\Gamma}$. Only when this contribution becomes deeper than the inertial one, $\bar{\Gamma}$ starts to strongly increase.

\section{Dynamo thresholds}
An important feature of the small-scale dynamo process is certainly the threshold Rm$\rm_c$, for which the amplification is activated.  We have evaluated Rm$\rm_c$ for all types of turbulence considered here. An asymptotic value of Rm$\rm_c$ = 320 has been found for  Kolmogorov turbulence and Re=$10^{14}$, which is in good agreement with other analytical \cite{Kleeorin2012} 
and numerical studies \cite{Schekochihin2005,Haugen2004,Malyshkin2010}, which report a Rm$\rm_c$ value of 410, 500, 210, and $\sim$500, respectively. Furthermore, we found that the threshold for the generation of magnetic fluctuations by highly compressible 
turbulent flows is considerably larger (Rm$\rm_c$ = 32000) than for the case of a Kolmogorov fluid. A similar trend was also shown in \cite{Schober2012,Rogachevskii1997}. 
It is worth noting that \citet{Leorat1981} obtained a critical magnetic Reynolds number for compressible Kolmogorov turbulence of the order of a few tens.

\begin{figure}
\includegraphics[width=.45\textwidth]{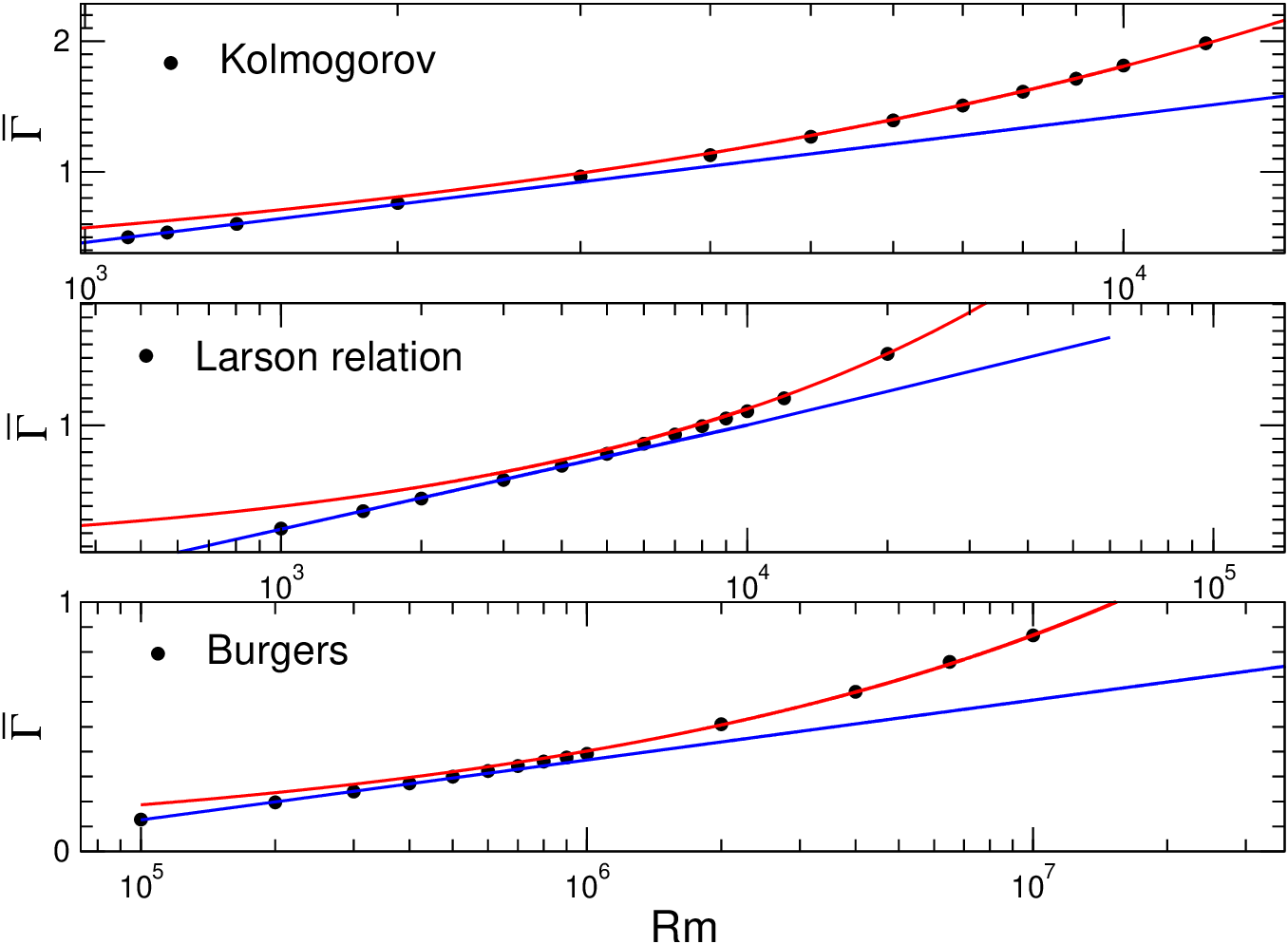}
\caption{Computed normalised growth rate $\bar{\Gamma}$ as a function of Rm for three different types of turbulence and for Pm~$\sim$~1.  Fitting curves are also reported in the panels. $\bar{\Gamma}$ scales as Rm$^\alpha$ (red line) when far from the threshold, and $\propto$ $\beta\ln({\rm Rm})$ in the vicinity of the threshold.}\label{figure4}
\end{figure}

In Fig. \ref{figure4}, we explore the dependence of the fastest growing mode on the magnetic Reynolds number for values in the vicinity of the threshold Rm$\rm_c$ for turbulence based on Kolmogorov, Burgers and the Larson relation. We adopt Pm $\simeq$ 1 and analyse the scaling of the 
growth rate in the vicinity and far away from the threshold (Rm$\rm_c$). For all cases considered here, the growth rate far from the threshold scales 
as $\propto$ Re$^{\alpha}$ for Pm $\gg$ 1 (or~$\propto$~Rm$^{\alpha}$ for Pm $\ll$ 1), with $\alpha$=1/2 for Kolmogorov and 1/3 for Burgers. In the vicinity of the threshold, the growth rate becomes a function of Re-Rm$\rm_c$. Here we perform a fit with a logarithmic function, $\bar{\Gamma}$ =  $\beta \ln({\rm Rm}) + \gamma$, with $\beta$ equal to $0.4214$, $0.3356$, and $0.10438$, and $\gamma$ = $\beta\ln(\rm Rm_c)$ to be -2.4516, -2.0901, and -1.0754 for Kolmogorov, Larson and Burgers turbulence, respectively \footnote{We note that the functional form near the threshold is not necessarily logarithmic, and a fit proportional to (Rm$^{1/2}$-Rm$_c^{1/2})$ also provides a valid description of the data (S. Boldyrev, private communication).}. Table \ref{table1} reports  the fitted growth rate as a function of Re and Rm for magnetic Reynolds numbers far from the threshold for the different types of turbulence, confirming the most efficient dynamo growth for the case of Kolmogorov turbulence.

\begin{table}
\caption{The normalised growth rate of the small-scale dynamo $\bar{\Gamma}$ as a function of Re and Rm for five different types of turbulence. The results for Pm $\ll$ 1, and Pm $\gg$ 1 are reported.}\label{table1}
\begin{tabular}{l|l|l}
\hline\hline
$\vartheta$ & Pm $\ll$ 1 & Pm  $\gg$ 1\\
\hline
1/3     &  1.85$\times10^{-2}$ Rm$^{1/2}$  &  9.98$\times10^{-1}$ Re$^{1/2}$  \\
0.38    &  1.80$\times10^{-2}$ Rm$^{0.45}$ &  7.62$\times10^{-1}$ Re$^{0.45}$ \\
0.43    &  1.31$\times10^{-2}$ Rm$^{0.40}$ &  5.09$\times10^{-1}$ Re$^{0.40}$ \\
0.47	&  8.91$\times10^{-3}$ Rm$^{0.36}$ &  3.07$\times10^{-1}$ Re$^{0.36}$ \\
1/2     &  3.69$\times10^{-3}$ Rm$^{1/3}$  &  1.54$\times10^{-1}$ Re$^{1/3}$  \\ 
\hline \hline
\end{tabular}
\end{table}

\section{Conclusions}
In this work we computed a numerical solution of the Kazantsev equation from the smallest to the largest magnetic Prandtl numbers for different types of turbulence with the aim of giving a general view of the full small-scale dynamo process. 
We discussed the features of the potential, which determine the behaviour of the growth rate. Different types of turbulence, from incompressible Kolmogorov to compressible Burgers, exploring the intermediate cases of 
turbulence observed in molecular clouds, have been analysed and the threshold for dynamo action been calculated. The growth rate of the magnetic energy decreases for more compressible types of turbulence, consistent with numerical simulations by Federrath et al. \cite{Federrath2011}. We stress the existence of dynamo action for Pm~$\ll$ 1 and confirm the earlier results reported in \cite{Schober2012} for Pm~$\rightarrow\infty$. We also included the range Pm~$\sim$ 1, where 
the often applied WKB approximation becomes invalid. The existence of higher growing modes contributing to the amplification of magnetic fields has been analysed for the whole range of Pm numbers confirming that a larger amplification is expected when it occurs in the viscous range. We find that for Kolmogorov turbulence, the growth rate scales as Re$^{1/2}$ for Pm~$\gg$~1, as Rm$^{1/2}$ for Pm~$\ll$~1 and as Re$^{1/3}$ and Rm$^{1/3}$ 
for Burgers-type turbulence. We further calculated the critical magnetic Reynolds numbers Rm$\rm_c$ for magnetic field amplification, which show a strong increase going from Kolmogorov to compressible Burgers turbulence, in good agreement with previous analytical and numerical results \cite{Kleeorin2012,Schekochihin2005,Haugen2004,Malyshkin2010}. 

A caveat that has to be kept in mind is the finite correlation time that is employed within the Kazantsev model. While many of its features, for instance the exponential growth and the dependence on the Reynolds numbers are in good agreement with numerical simulations, it may nevertheless influence its properties to some extend. For instance, \citet{Scheko01} calculated the first order corrections of the dynamo growth rate due to the finite correlation time, suggesting an overall reduction of the growth rate by up to $40\%$. An additional feature that may result from finite correlation times is the so-called Golitsyn spectrum \citep{Kleeorin2012}, which appears as an additional term in the correlation function of the magnetic field, although it never becomes dominant. Beyond the techniques employed in their studies, only the comparison with numerical simulations will shed further light on the effect of finite correlation times \citep{Plunian2010,Sahoo2011,Ponty2005, Federrath2010, Federrath2011}.

Our results thus illustrate the behavior of the small-scale dynamo under a larger range of astrophysically relevant conditions, including the interior of stars, the interstellar medium, the intergalactic medium and the first stars and galaxies. As originally suggested by \cite{Beck96}, the initial phase of magnetic field amplification via the small-scale dynamo is crucial for providing strong seeds on which the $\alpha$-$\omega$ dynamo can subsequently act. Indeed, high-resolution numerical simulations find evidence for turbulence already in the first protogalaxies, suggesting that the dynamo works early on \cite{Latif2012}. Due to the efficiency of the dynamo even in the highly compressible regime, the scenario seems capable of explaining the magnetic field structures in present-day galaxies.

\begin{acknowledgments}
We thank for funding through the DFG priority  program 'The Physics of the interstellar medium' (projects SCHL 1964/1-1 and KL 1358/14-1). D.R.G.S. acknowledges for funding via
the SFB 963/1 on 'Astrophysical flow instabilities and turbulence'. J. S. thanks IMPRS HD. We thank the anonymous referees for valuable comments on our manuscript.
\end{acknowledgments}
\bibliography{MagneticFields}

\end{document}